\documentstyle[emulateapj,psfig]{article}
 
\slugcomment{Received 1999 June 16; accepted 1999 July 7}

\begin{document}

\title{{\it HST} Observations of the Serendipitous X-ray Companion to 
M\lowercase{rk}~273:  Cluster at \lowercase{z} = 0.46?\footnote{
Based on observations with the NASA/ESA Hubble Space Telescope, obtained at
the Space Telescope Science Institute, which is operated by the Association of
Universities for Research in Astronomy, Inc. under NASA contract No.
NAS5-26555.}
}

\author{Kirk D. Borne\altaffilmark{2}}
\affil{Raytheon Information Technology and Scientific Services,
NASA Goddard Space Flight Center, Greenbelt, MD 20771 \\
{\tt{borne@rings.gsfc.nasa.gov}}}
 
\author{Luis Colina}
\affil{Instituto de Fisica de Cantabria (CSIC-UC),
Facultad de Ciencias, 39005 Santander, Spain}
 
\and
 
\author{Howard Bushouse and Ray A. Lucas}
\affil{Space Telescope Science Institute, 3700 San Martin Drive,
Baltimore, MD 21218}

\altaffiltext{2}{Sabbatical Visitor, Space Telescope Science Institute}

\begin{abstract}

We have used HST $I$-band images to identify Mrk~273X, the very unusual
high-redshift X-ray-luminous Seyfert~2 galaxy found by ROSAT in the
same field-of-view as Mrk~273.  We have measured the photometric
properties of Mrk~273X and have also analyzed the luminosity
distribution of the faint galaxy population seen in the HST image.  The
luminosity of the galaxy and the properties of the surrounding
environment suggest that Mrk~273X is the brightest galaxy in a
relatively poor cluster at $z \approx 0.46$.  Its off-center location
in the cluster and the presence of other galaxy groupings in the HST
image may indicate that this is a dynamically young cluster on the
verge of merging with its neighboring clusters.  We find that Mrk~273X
is a bright featureless elliptical galaxy with no evidence for
a disk.  It follows the de~Vaucouleurs ($r^{1/4}$) surface brightness law
very well over a range of 8 magnitudes.  Though the surface brightness 
profile does not appear to be dominated by the AGN, the galaxy has very 
blue colors that do appear to be produced by the AGN.  Mrk~273X is most 
similar to the IC~5063 class of active galaxies --- a hybrid 
Sy~2 / powerful radio galaxy.

\end{abstract}


\section{Introduction} \label{introd}

Mrk 273 (IRAS 13428+5608) is one of the nearest members (at $z$=0.0378)
of a special class of galaxies --- the ultraluminous infrared galaxies
(ULIRGs) --- identified by IRAS (Sanders {\it{et al.}}~1988a,b).  It
shows a long tidal tail and a disturbed morphology, thus indicating its
presumed galaxy-galaxy collision+merger origin.  The merger and its
accompanying starburst and/or AGN activity leads to the high IR
luminosity ($> 10^{12} L_\odot$) through dust absorption and
re-emission of the intense but obscured radiation field.  Mrk~273 has a
Sy~2 nuclear spectrum (Lutz, Veilleux, \& Genzel 1999) and
consequently is a soft X-ray source (Turner {\it{et al.}}~1993).  ROSAT
PSPC X-ray observations of the field surrounding this galaxy and
several other Sy~2 galaxies revealed the presence of serendipitous
companion X-ray sources around each primary source (Turner {\it{et
al.}}~1993; see also Laurikainen \& Salo 1995, Radecke 1997, Arp 1997).
The X-ray companion to Mrk~273 (hereafter, Mrk~273X) is
1.3$'$ to the northeast (projected separation $\approx$ 57~kpc, at the
redshift of Mrk~273), with a soft X-ray count rate $\sim$50\% that of
Mrk~273.  (For this paper, we assume $H_0$ = 70 km s$^{-1}$ Mpc$^{-1}$
and $q_o = {1 \over 2}$.)

Recent spectroscopic observations of Mrk~273X have revealed that it is
a Sy~2 galaxy itself, though at a much higher redshift ($z$=0.458) and
with some very unusual properties (Xia {\it{et al.}}~1998a,b,
1999).  Xia {\it{et al.}}~(1999) report a soft X-ray luminosity 
($6.1 \times 10^{43}$ erg s$^{-1}$) that is extraordinarily
high compared to most Sy~2 galaxies even though the photon power law
index has a typical Sy~2 value (-1.98).  The radio luminosity
($L_{\rm{1.37Ghz}} \approx 1.0 \times 10^{40}$ erg s$^{-1}$) is also
quite high for a Sy~2 (Yun \& Hibbard 1999).  Furthermore, the derived
column density of neutral hydrogen (from the X-ray spectrum) is very
low for a Sy~2:  $\log(N_H) \approx 20.6$.  Many Sy~2 galaxies have
$\log(N_H) > 21$ (Turner {\it{et al.}}~1997), though values similar to
Mrk~273X are not uncommon (Turner {\it{et al.}}~1998).  High X-ray and
radio power along with low $N_H$ are more indicative of an unobscured
view of the AGN (i.e., as in a Sy~1), and thus these observations
challenge the conventional inclined dusty torus models for AGN.  
Xia {\it{et al.}}~(1999) thereby contrast the possibilities that Mrk~273X 
may be either a Sy~2 or a narrow-line Sy~1 (NLS1) galaxy.  They conclude
that various optical emission line ratios and the X-ray spectral index
weigh strongly against the NLS1 hypothesis and in favor of the Sy~2
interpretation.  Xia {\it{et al.}}~(1999) additionally discuss the
evidence for X-ray variability in Mrk~273X, which is still not certain
but nevertheless possible.

Given the unusual properties of this faint galaxy, we were pleased to
find it within the field-of-fiew of our HST image of Mrk~273 --- part
of our large HST survey of ULIRGs (Borne {\it{et al.}}~1997a,b,
1999a,b,c).  We report here on our analysis of the Mrk~273X image and
of its surrounding field (which includes many comparably faint
galaxies).  We describe the HST observations in \S 2 and the results of
our image analysis in \S 3.  The latter includes photometric and
luminosity function derivations. We also summarize and discuss in \S 3
the properties of this galaxy in relation to the properties of
analogous active galaxies.

\section{HST Imaging Observations}

We obtained two sets of HST imaging observations of the primary ULIRG
target, Mrk~273, as part of our large ULIRG Snapshot Survey program.
In each set of observations, we used the WFPC2 camera to obtain two
400-sec images in the I-band (F814W filter).  Pairs of images were used
to remove the effects of cosmic-ray radiation events in the CCDs
(assuming the events are uncorrelated and have no persistence from one
image to the next).  In one set of observations, Mrk~273 was centered
in the WF3 CCD (with 0.10$''$ per pixel), and in a second set, Mrk~273
was centered in the PC CCD (0.046$''$ per pixel).  In the latter
observations, we found the X-ray companion Mrk~273X in the WF4 CCD
frame (Fig.~\ref{fig:gals}).  The standard calibrated data products
were re-derived using the best calibration files and then were combined
into a single cosmic-ray cleaned 800-sec image.  Our analysis was
carried out using this final cleaned image.

\section{Results of Image Analysis}

\subsection{Identification of Mrk~273X} \label{gal-id}

From their ROSAT PSPC X-ray images, Turner {\it{et al.}}~(1993) 
give a boresight-corrected J2000 position for Mrk~273X at
13$^h$44$^m$47.9$^s$, +55$^o$54$'$11$''$.  Using this information and
the images in Xia {\it{et al.}}~(1998a, 1999), we have identified the
optical counterpart in our WFPC2 image.  We find the J2000 position
(within the standard uncertainty of $\pm 0.5''$ for HST positions) for
Mrk~273X to be: 13$^h$44$^m$47.46$^s$, +55$^o$54$'$11.1$''$.  This is almost
exactly the position given by Turner {\it{et al.}}~(1993).  We show in
Figure~\ref{fig:gals} the $75'' \times 75''$ usable area of the WF4 CCD
frame containing Mrk~273X.  About 40 fainter galaxies are also seen in
this frame.  The darker grey shading of the background light
distribution in the lower right quadrant (southwest) is real --- this
is the extended spray of emission from the tidal debris around Mrk~273,
as indicated by Xia {\it{et al.}}~(1998a, 1999).  Mrk~273 is located 
in the PC frame just below the lower right quadrant of the WF4 frame 
shown here.

Mrk~273X is well resolved, with measurable light out to a radius of
$\sim$50 pixels (5.0$'' \approx$ 25 kpc).  It is an essentially 
featureless galaxy, probably an elliptical or S0, with a small 
degree of flattening.  It has a bright nucleus, typical of an 
elliptical galaxy.

\subsection{Photometric Properties of Mrk~273X} \label{photprops}

\begin{figure} 
\par\noindent
\centerline{\psfig{figure=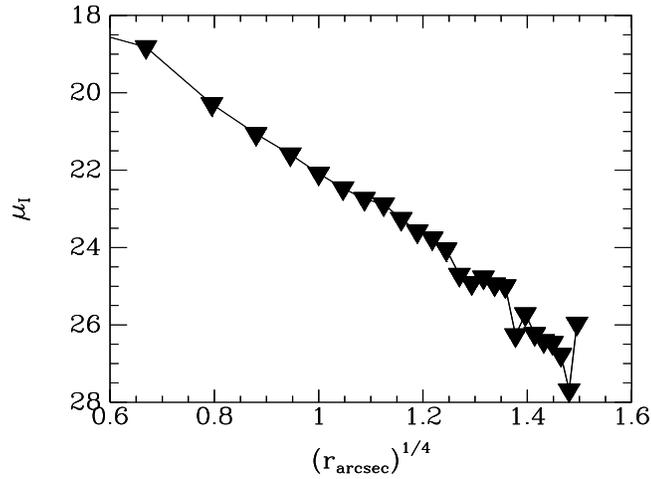,height=6.68cm,width=9cm,angle=-90}}
\caption{
Derived intensity profile for Mrk~273X from the IRAF/STSDAS ELLIPSE task.
The intensity scale is in units of $I$ magnitude per square
arcsec, and the radial scale
is in (arcsec)$^{1 \over 4}$ units.  Note that the profile follows
nearly a straight line (de Vaucouleurs profile), as would a typical elliptical 
galaxy.  A very tiny (almost imperceptable) rise in the curve at 
$r^{1/4} \approx 1.12$ may be caused by the very small companion 
seen embedded in the optical halo of Mrk~273X (Fig.~\ref{fig:mosaic},
at 1.6$''$ east).  We did not mask out that companion in our ELLIPSE
fit due to its very low brightness --- we see here that the effect
of including it is minimal.
}
\label{fig:devauc}
\end{figure}

\begin{figure} 
\par\noindent
\centerline{\psfig{figure=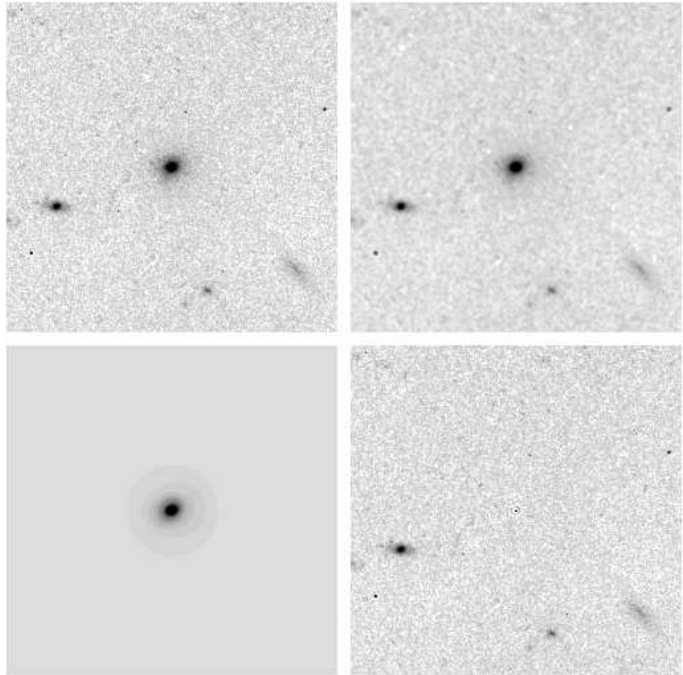,height=9cm,width=9cm}}
\caption{
{\it{Upper left}}: 25$''\times$25$''$ section of the HST WFPC2 image, 
centered on Mrk~273X (where 25$''$ corresponds to 125 kpc).  
At least three (and maybe as many as five)
companion galaxies are seen within $\sim$10$''$ ($\approx$ 50 kpc).
North is up and east is to the left.  Note the smooth elliptical
appearance of Mrk~273X, with no evidence for a disk or flattened component.
{\it{Upper right}}: 3$\times$3-pixel boxcar-smoothed version of HST image.
{\it{Lower left}}: IRAF/STSDAS ELLIPSE model for Mrk~273X (fit
out to radius = 5$''$).  The mean
ellipticity of the model (out to radius = 1$''$) is 0.17, 
and the mean isophotal position angle is 115$^\circ$.
The half-light radius is 0.4$''$ (= 2.0 kpc).
{\it{Lower right}}: Residual HST image, produced by subtracting the model
({\it{lower left}}) for Mrk~273X from the upper left image.  Note the
near perfect subtraction of the Mrk~273X galaxy image (except at
the very center), indicating the
very good approximation of the ELLIPSE-fit model to the actual data.
}
\label{fig:mosaic}
\end{figure}

We used various photometric tasks within IRAF and STSDAS to analyze
Mrk~273X and its surrounding galaxies.  The ELLIPSE task was used to
analyze the surface brightness profile and shape of Mrk~273X.  We find
that the half-light radius for Mrk~273X is 0.40$''$ = 2.0 kpc and the
radially averaged ellipticity of the galaxy corresponds approximately
to an E2 shape, at an average position angle of $\sim$115$^\circ$.  
The derived intensity
profile is shown in Figure~\ref{fig:devauc}.  We find that the light
follows the expected $r^{1 \over 4}$ profile for an
elliptical galaxy very well --- over a range of 8 magnitudes
in surface brightness.  
To determine how well the data are matched by the derived
ELLIPSE model, we constructed a model galaxy using the derived
parameters and subtracted that model from the data to produce a
residual map.  The different steps in this process are illustrated in
Figure~\ref{fig:mosaic}, including the original data ({\it{upper
left}}), a 3$\times$3-pixel boxcar-smoothed version of the data
({\it{upper right}}), the constructed model ({\it{lower left}}), and
the residual (data minus model) image ({\it{lower right}}).  From the
near-zero residuals, we see that our ``elliptical galaxy'' model
matches the data quite well.  

According to Graham \& Colless (1997),
$R_{\rm{eff}} = 0.75 R_{1 \over 2}$ for a wide
range of luminosity profile shapes and galaxy models. 
Assuming that this applies to Mrk~273X, we find
$R_{\rm{eff}} = 0.30''$ = 1.5 kpc, which
is small compared to most ellipticals
(Scodeggio, Giovanelli, \& Haynes 1998).
 
A very small companion galaxy is seen in the optical halo of the
galaxy, approximately 16 pixels (1.6$'' \approx$ 8~kpc) east of 
Mrk~273X's center (see upper panels in Fig.~\ref{fig:mosaic}).  
In addition, three brighter companions are seen within 10$''$
($\approx$ 50~kpc) of Mrk~273X, and possibly two additional
fainter companions are also seen within that distance (south-southwest)
of Mrk~273X), as shown in Figure~\ref{fig:mosaic}.
Note that the brightest of the companion galaxies have
asymmetric and disk-like morphologies that are clearly evident at this
spatial resolution (0.2$''$ = 1~kpc) and low $S/N$.  Such features are
not seen at all in Mrk~273X, which therefore clearly possesses an
ellipsoidal early-type galaxy morphology.

The ELLIPSE model does show significant isophote twisting ($\Delta$PA
$\approx 30^\circ$) within the central 2$''$ ($\sim$10~kpc).  This may
be induced in Mrk~273X through interactions with the close companions.
If the companions have that degree of influence on Mrk~273X, then their
interaction may also be responsible for tidally triggering the AGN
activity.

We used the IRAF APPHOT task to measure the I-band magnitude of
Mrk~273X and of the surrounding galaxies (see \S \ref{lumfun}).  The
NASA Extragalactic Database (NED) gives a magnitude of 19.6 (no
passband specified, but probably R-band) for Mrk~273X.  We find $m_I =
19.10$ (Cousins I) for the total light.  For $z=0.458$: $m-M = 41.66$
and therefore $M_I = -22.56$, which corresponds to the rest-wavelength
$V$-band.

Xia {\it{et al.}}~(1998a, 1999) report B=20.8 and R=19.6.  Therefore,
the colors of Mrk~273X are: $B-R$=1.2 and $B-I$=1.7.  We compare these
colors in Table~\ref{tab:colors} with those measured for the galaxies
in the $z$=0.41 cluster Cl~0939+472 studied by Belloni \& Roser
(1996).  The distance of that cluster is similar to that of Mrk~273X
and so the K-correction can be ignored in the comparison of colors.  We
find that the $B-R$ and $B-I$ colors of Mrk~273X are consistent with
the colors of the late-type Im and Scd cluster galaxies, both in the
mean value and in the observed ranges of these colors.

\subsection{The Effect of an AGN} \label{agn-effects}

Based on the color information (Table~\ref{tab:colors}), we conclude
that Mrk~273X is either a very late type galaxy (i.e., with recent star
formation) or else its colors are seriously affected by the AGN (Sy~2
nucleus), or both.  On-going star formation in Mrk~273X would suggest
the presence of gas, which would provide a source of fuel for the AGN,
and the nearby companions (Fig.~\ref{fig:mosaic}) could provide a
possible trigger for the activity.

As we see in Figure~\ref{fig:devauc}, the surface brightness
of the galaxy follows the standard elliptical galaxy radial variation,
providing no photometric evidence for an AGN point source
contaminating the core brightness profile.  However, the small measured
value for the effective radius (1.5 kpc) may be an effect of the AGN 
contributing some fraction of the light in the core.

Mrk~273X has the optical spectral properties of a Sy~2 galaxy, but the
radio flux, soft X-ray flux, optical morphology, and cluster dominance
(\S \ref{lumfun}) of a powerful radio galaxy (PRG).  We compare and
contrast various of these properties with the properties of analogous
galaxies in Table~\ref{tab:agn}.  We see there that the properties of 
Mrk~273X span the range of the different types of active galaxies and yet
do not correspond to any one AGN type.  Its properties are atypical for Sy~2
galaxies in that Mrk~273X has very high $L_{opt}$, $L_{radio}$, $L_{SX}$,
and $L_{H\alpha}$, but very low $N_H$ (see \S \ref{introd}).  Its
properties are most similar to IC~5063 --- only $L_{SX}$ differs
significantly between the two sources.  We know that IC~5063 has a very
high column density ($\log N_H = 23.3$) and a high hard X-ray
luminosity ($\log L_{HX} = 43.04$), as measured by Koyama 
{\it{et al.}}~(1992).  Thus the total (soft$+$hard) X-ray luminosity of 
the two sources is nearly the same (as are the optical, radio, and
H$\alpha$ luminosities).  Mrk~273X is therefore a galaxy of 
the IC~5063-type, except that its low $N_H$ allows a high flux of soft
X-rays to escape.  Inglis {\it{et al.}}~(1993) found that IC~5063 shows
broad lines in polarized light and thus likely contains an obscured PRG
or Sy~1 nucleus (Morganti {\it{et al.}}~1998).  Based on
these comparisons, we believe that Mrk~273X is also a PRG.

\subsection{Luminosity Function of Surrounding Galaxies} \label{lumfun}

The brightest of the other galaxies seen in the WF4 frame are fainter
than Mrk~273X, but they are all comparably bright.  We have examined
the full WFPC2 image (including the WF2 and WF3 frames) and we note
that there are a couple of other small groupings of galaxies of similar
brightness, number count, and spatial extent.  However, those other
groupings are relatively far from the WF4 frame (e.g., at the far edge
of the adjacent WF3 frame and on the far half of the diagonally
opposite WF2 frame) --- at distances of 75$''$, 110$''$, and 130$''$
(where 100$''$ = 490~kpc).  Even though they are not spatially close to
the galaxies seen in WF4, these other groups could be associated with
Mrk~273X nevertheless given that all of these galaxies have similar
sizes and apparent magnitudes (i.e., at a similar redshift).  To first
order, given the observed spatial segregation of these groupings, we
believe that the galaxies seen in WF4 comprise a small isolated group,
of which Mrk~273X is the brightest member (at least, it is the brightest
member that we have available within our WF4 image).  Its rest-wavelength
$V$ absolute magnitude (\S \ref{photprops}) is consistent with 
this being a BCG (brightest cluster galaxy; Postman \& Lauer 1995).
The BCG status is further supported by the $R$ magnitude (= 19.6), which 
makes Mrk~273X comparable in brightness to the brightest galaxies 
(ellipticals) in the $z$=0.41 cluster CL~0939+4713 that was studied
with HST by Dressler {\it{et al.}}~(1994a,b).

\begin{figure} 
\par\noindent
\centerline{\psfig{figure=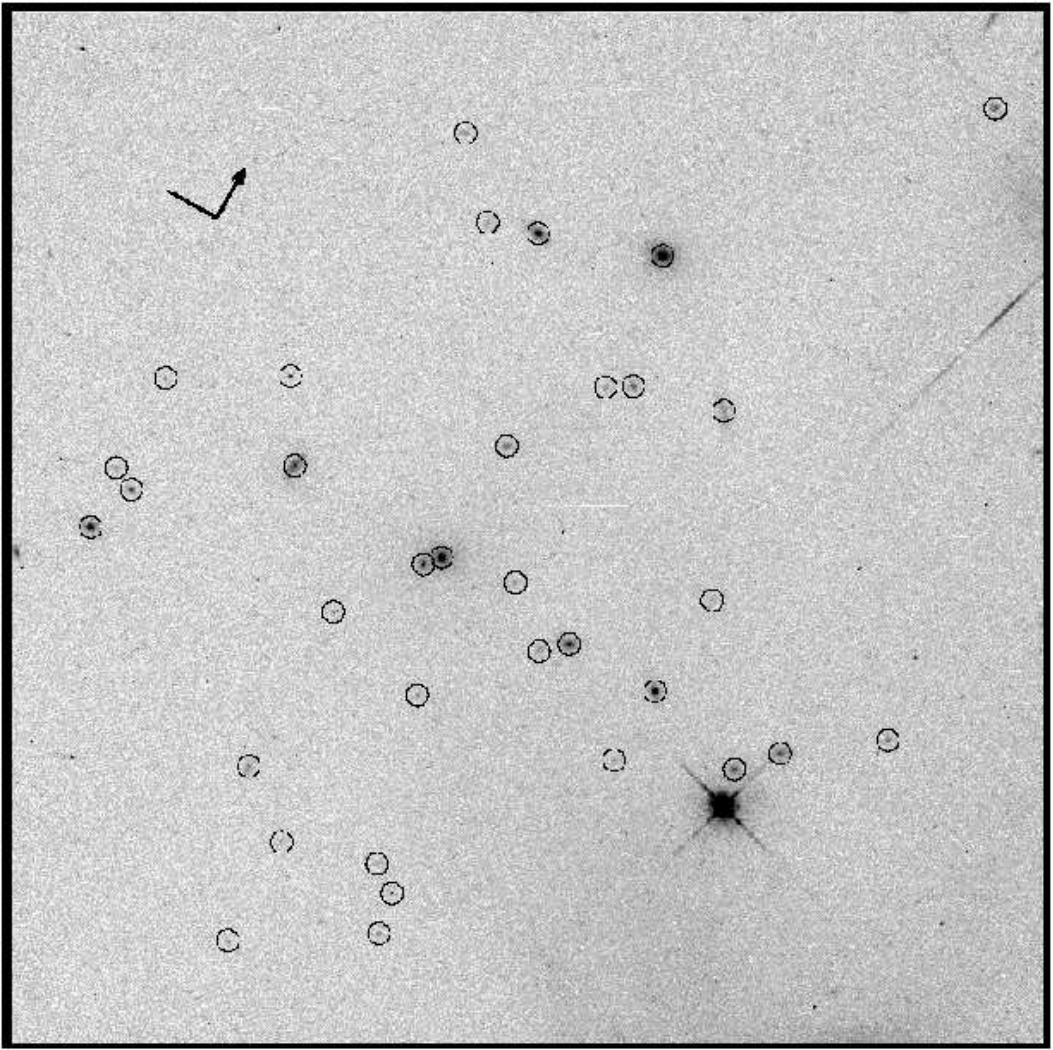,height=5.5in,width=5.5in}}
\caption{
75$''\times$75$''$ section of HST WFPC2 image, showing the usable
portion of the WF4 CCD frame.  At the redshift of Mrk~273X ($z$=0.458),
75$''$ corresponds to 370 kpc.  The arrow on the North-East indicator
points north.  All of the galaxies whose estimated $I$ magnitude is
brighter than 24.0 are circled.  These are used to produce the
luminosity function shown in the top panel of Fig.~\ref{fig:histograms}
--- the magnitudes represented in Fig.~\ref{fig:histograms} were
measured within a circular aperture of radius 0.8$''$ (= 8 pixels),
corresponding to the circles around each marked galaxy in this image.
The brightest galaxy in the frame is Mrk~273X --- it is the brightest
object in the upper right quadrant of the image.  The dumbbell galaxy
pair just to the left of center has a combined I magnitude of 19.5,
still fainter than the $m_I$=19.1 for Mrk~273X.  There are 34 galaxies
marked with $m_I < 24.0$.  At the redshift of Mrk~273X, $m-M = 41.66$,
so that the faintest galaxy marked has $M_I \approx -17.7$ (if it is at
the same distance as Mrk~273X).
}
\label{fig:gals}
\end{figure}

\begin{figure} 
\par\noindent
\centerline{\psfig{figure=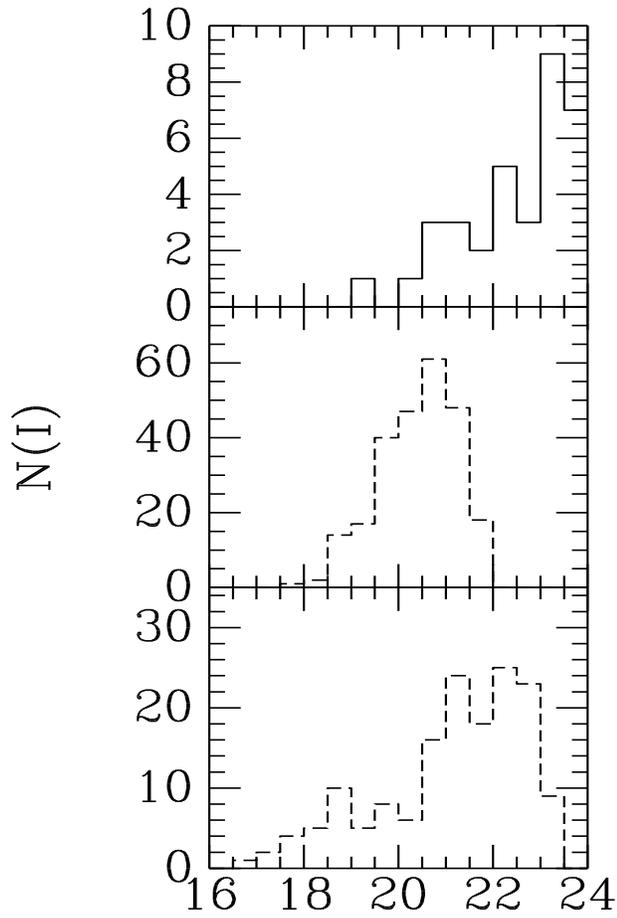,height=4.90in,width=6.60in,angle=-90}}
\caption{
Luminosity function for 3 regions.  Each histogram shows
the number distribution of galaxies as a function of I magnitude
(Gunn $i$ for the bottom plot).
Top: field around Mrk~273X (z=0.458; this paper).  
Middle: cluster CL~0939+472 (z=0.41; Belloni \& Roser 1996).
Bottom: cluster CL~2158+0351 (z=0.45; Molinari {\it{et al.}}~1990).  
The galaxies measured for
the Mrk~273X distribution (top panel) are those that are 
indicated in Fig.~\ref{fig:gals}.
Both Belloni \& Roser (1996) and Molinari {\it{et al.}}~(1990)
had magnitude cutoffs in their samples: $R=22.5$ and Gunn $r=23.5$,
respectively.
}
\label{fig:histograms}
\end{figure}

We used the IRAF APPHOT task to measure a metric $I$-band magnitude for
the 34 circled galaxies in Figure~\ref{fig:gals}.  We measured the flux
within a radius of 8 pixels (0.8$''$ = 4 kpc) and included only those
galaxies with $I$-band metric magnitude brighter than 24.0.  (Fainter galaxies
could not be measured reliably in this short-exposure image.)  Within
our fixed metric aperture, Mrk~273X has $I$=19.5 (compared with $I$=19.1
for its total light).  The spatial distribution of the marked galaxies in
Figure~\ref{fig:gals} shows that Mrk~273X is far ($\sim$200 kpc)
from the center of the group.  In fact, a bright dumbbell pair of 
galaxies is seen near the center, but the pair's combined {\it{total light}} 
has $I$=19.5.  We show in Figure~\ref{fig:histograms} the luminosity 
function for the 34 galaxies in our WF4 frame along with the $I$-band 
luminosity function for two clusters of galaxies at nearly the same 
redshift:  cluster CL~0939+472 (z=0.41) from Belloni \& Roser (1996), 
and cluster CL~2158+0351 (z=0.45) from Molinari {\it{et al.}}~(1990).

A comparison of the histograms in Figure~\ref{fig:histograms} reveals
that the luminosity distribution of galaxies in the field surrounding
Mrk~273X is similar to the bright end of a typical cluster luminosity
function at that redshift.  This further supports the notion that
Mrk~273X is the brightest member of a poor cluster of galaxies at $z$=0.458.
Given this galaxy's non-central location within
the group, this is probably a dynamically young still-evolving cluster,
perhaps still collapsing.  In fact, this group may be on the verge of
merging with the other small groups of galaxies seen in our wider WFPC2
field-of-view (see above).  We note that there was no
evidence in the X-ray images for an extended cluster-like hot ICM
within this group.

\section{Summary}

We have analyzed HST images of Mrk~273X, the serendipitously discovered
X-ray companion to Mrk~273.  Mrk~273X is at a much higher redshift and
therefore not physically associated with Mrk~273 (Xia {\it{et al.}}~1999).
Mrk~273X is a featureless early type galaxy and appears to be the
brightest member of a small cluster of galaxies.  The optical
morphology of Mrk~273X (including its radial surface brightness profile)
and its role as the dominant member of a 
cluster resemble the properties of a PRG:  an elliptical or
other early type galaxy.  However, its colors and Sy~2 spectrum are
typical of much later galaxy types.  This suggests that the galaxy's
colors are strongly contaminated by the AGN (through
both its blue continuum and its
emission lines).  We believe that Mrk~273X is an active galaxy of the
IC~5063 type, except that the soft X-ray source in Mrk~273X is not
obscured as it is in IC~5063.  Mrk~273X therefore appears to be a
selectively obscured PRG in that the radio core and X-ray emitting
region are exposed (as in a typical PRG or Sy~1), but the broad
line-emitting region is obscured (as in a typical Sy~2).  This 
may indicate that the obscuring torus has an intermediate 
line-of-sight inclination.
Followup observations (particularly, redshift determinations) of the
galaxies surrounding Mrk~273X would validate the cluster hypothesis and
thus shed some light on the dynamical state of this system and possibly
lead to an identification of the trigger (tidal companion) for the AGN
activity.

\acknowledgements

Support for this work was provided by NASA through 
grant number GO-06346.01-95A from the Space
Telescope Science Institute, which is operated by 
AURA, Inc., under NASA contract NAS5-26555.
KB thanks Raytheon for providing financial support
during his Sabbatical Leave.  KB also thanks Ron Allen
and the Space Telescope Science Institute for their hospitality
in sponsoring his Sabbatical Visit.
This research has made use of the NASA/IPAC Extragalactic Database (NED) 
which is operated by the Jet Propulsion Laboratory, California Institute 
of Technology, under contract with the National Aeronautics and Space 
Administration.
This research has also made use of the 
on-line astronomical data services 
({\tt{http://adc.gsfc.nasa.gov/}})
at the Astronomical Data Center
(ADC) at NASA Goddard Space Flight Center.

\newpage

\newpage

\begin{deluxetable}{cccccc}
\tablewidth{450pt}
\small
\tablecaption{Colors of Mrk~273X vs. Cluster Galaxies at $z$=0.41
(Belloni \& Roser 1996) \label{tab:colors} }
\tablehead{
\colhead{Galaxy Type} & 
\colhead{Number} & 
\colhead{$<B-R>$} & 
\colhead{Range of $B-R$} & 
\colhead{$<B-I>$} & 
\colhead{Range of $B-I$} 
}
\startdata
E           & 139    & 2.4     & 1.6--3.0       & 3.5     & 2.5--4.4 \nl
E+A         & 42     & 2.0     & 1.4--2.5       & 3.0     & 2.3--3.7 \nl
Sbc         & 32     & 1.7     & 1.3--2.1       & 2.6     & 2.0--3.2 \nl
Scd         & 16     & 1.4     & 0.8--2.4       & 2.2     & 1.7--3.4 \nl
Im          & 17     & 1.1     & 0.9--1.4       & 1.7     & 1.5--2.0 \nl
Mrk~273X & ... &\multicolumn{2}{c}{$B-R=1.2$} &\multicolumn{2}{c}{$B-I=1.7$} \nl
\enddata
\end{deluxetable}

\begin{deluxetable}{lcccccccc}
\tablewidth{450pt}
\tablecaption{Properties of Mrk~273X and Comparison Active Galaxies
\label{tab:agn} }
\scriptsize
\tablehead{
\colhead{Galaxy} & 
\colhead{Type} & 
\colhead{$z$} & 
\colhead{Distance} & 
\colhead{$I$} & 
\colhead{$M_I$} & 
\colhead{$\log(L_X^{\rm soft})$} & 
\colhead{$\log(L_{\rm radio}^{\rm 1.4 GHz})$} & 
\colhead{$\log(L_{H\alpha})$} \\
\colhead{}   & 
\colhead{}   & 
\colhead{}   & 
\colhead{(Mpc)} & 
\colhead{(mag)} & 
\colhead{(mag)} & 
\colhead{(erg/s)} & 
\colhead{(erg/s)} & 
\colhead{(erg/s)} 
}
\startdata 
NGC 5506  & Sy 2      & 0.00618 & 26.5 
& 13.12\tablenotemark{a}
& -18.9 
& 41.98\tablenotemark{b}
& 38.57\tablenotemark{c}
& 40.51\tablenotemark{d} \nl
NGC 2992  & Sy 2      & 0.00771 & 33.0 
& 12.42\tablenotemark{a}
& -20.1 
& 42.33\tablenotemark{b}
& 38.52\tablenotemark{e}
& 40.90\tablenotemark{f} \nl
IC 5063   & Sy 2 / RG & 0.01135 & 48.7 
& 10.64\tablenotemark{g}
& -22.8 
& $<$41.40\tablenotemark{b,h}
& 39.60\tablenotemark{i}
& 41.23\tablenotemark{j} \nl
Fairall 9 & Sy 1      & 0.04702 & 204 
& 13.16\tablenotemark{a}
& -23.3 
& 44.10\tablenotemark{k}
&   ? 
& 43.48\tablenotemark{l} \nl
3C 273    & QSO       & 0.15834 & 703 
& 12.17\tablenotemark{a}
& -27.1 
& 45.8\tablenotemark{m}
& 43.62\tablenotemark{c}
& 44.81\tablenotemark{l} \nl
Mrk~273X  & Sy 2      & 0.458   & 2150 
& 19.10\tablenotemark{n}
& -22.6 
& 43.8\tablenotemark{o}
& 40.0\tablenotemark{p}
& 41.58\tablenotemark{q} \nl
\enddata
\tablenotetext{a}{Kotilainen et al.~(1993)}
\tablenotetext{b}{Fabbiano et al.~(1992)}
\tablenotetext{c}{White \& Becker~(1992)}
\tablenotetext{d}{Storchi-Bergman et al.~(1995)}
\tablenotetext{e}{Ulvestad \& Wilson~(1984)}
\tablenotetext{f}{Forbes \& Ward~(1993)}
\tablenotetext{g}{Estimated from $V-I=1.23$ (Buta \& Williams 1995) 
and $V=11.87$ (RC3)}
\tablenotetext{h}{Koyama et al.~(1992) measured $\log N_H = 23.3$ 
and the hard X-ray flux, yielding $\log L_{HX} = 43.04$}
\tablenotetext{i}{Morganti et al.~(1998)}
\tablenotetext{j}{Colina et al.~(1991)}
\tablenotetext{k}{Ceballos \& Barcons (1996)}
\tablenotetext{l}{Steiner (1981)}
\tablenotetext{m}{Brinkmann et al.~(1994)}
\tablenotetext{n}{This paper}
\tablenotetext{o}{Turner et al.~(1993)}
\tablenotetext{p}{Yun \& Hibbard~(1999)}
\tablenotetext{q}{Xia et al.~(1999)}
\end{deluxetable}


\begin{thebibliography}{}


\bibitem{arp97}
Arp, H.
1997, A\&A, 319, 33

\bibitem{belloni96}
Belloni, P., \& Roser, H.-J. 
1996, A\&ASS, 118, 65

\bibitem{borne97a}
Borne, K., Bushouse, H., Colina, L., \& Lucas, R. A.
1997a, Revista Mexicana de Astronomia y Astrofisica Serie de Conferencias, 
6, 250

\bibitem{borne97b}
Borne, K., Bushouse, H., Colina, L., \& Lucas, R. A.
1997b, Extragalactic Astronomy in the Infrared, 
eds. G. A. Mamon~et al. (Paris: Editions Frontieres), 277
 
\bibitem{borne99a}
Borne, K., Bushouse, H., Colina, L., \& Lucas, R. A.
1999a, After the Dark Ages: When Galaxies were Young,
eds. S. Holt \& E. Smith (Woodbury: AIP), 220
 
\bibitem{borne99b}
Borne, K., Bushouse, H., Colina, L., Lucas, R. A., Baker, A.,
Clements, D., Lawrence, A., \& Rowan-Robinson, M.
1999b, Astrophysics with Infrared Surveys: A Prelude to SIRTF, 
PASP Conference Proceedings, in press (astro-ph/9809040)

\bibitem{borne99c}
Borne, K., Bushouse, H., Colina, L., Lucas, R. A., Baker, A.,
Clements, D., Lawrence, A., \& Rowan-Robinson, M.
1999c, ApSS, in press (astro-ph/9902293)

\bibitem{buta95}
Buta, R., \& Williams, K. L.
1995, AJ, 109, 543

\bibitem{brinkmann94}
Brinkmann, W., Siebert, J., \& Boller, T.
1994, A\&A, 281, 355

\bibitem{ceballos96}
Ceballos, M. T., \& Barcons, X. 
1996, MNRAS, 282, 493

\bibitem{colina91}
Colina, L., Sparks, W. B., \& Macchetto, F. 
1991, ApJ, 370, 102

\bibitem{RC3}
de Vaucouleurs, G., de Vaucouleurs, A., Corwin Jr., H. G., Buta, R. J., 
Paturel, G., \& Fouque, P. 
1991, Third Reference Catalogue of Bright Galaxies (Springer: New York) (RC3)

\bibitem{dressler94a}
Dressler, A., Oemler, Jr., A., Butcher, H. R., \& Gunn, J. E.
1994a, ApJ, 430, 107

\bibitem{dressler94b}
Dressler, A., Oemler, Jr., A., Sparks, W. B., \& Lucas, R. A.
1994b, ApJ, 435, L23

\bibitem{fabbiano92}
Fabbiano, G., Kim. D. W., \& Trinchieri, G. 
1992, ApJS, 80, 531

\bibitem{forbes93}
Forbes, D., \& Ward, M. J. 
1993, ApJ, 416, 150

\bibitem{graham97} 
Graham, A., \& Colless, M. 
1997, MNRAS, 287, 221
 
\bibitem{inglis93}
Inglis, M. D., Brindle, C., Hough, J. H., Young, S., Axon, D. J., 
Bailey, J. A., \& Ward, M. J. 1993, MNRAS, 263, 895

\bibitem{kotilainen93}
Kotilainen, J. K., Ward, M. J., \& Williger, G. M. 
1993, MNRAS, 263, 655

\bibitem{koyama92}
Koyama, K., Awaki, H., Iwasawa, K., \& Ward, M. 
1992, ApJ, 399, L129

\bibitem{laurik95}
Laurikainen, E., \& Salo, H.
1995, A\&A, 293, 683

\bibitem{lutz99}
Lutz, D., Veilleux, S., \& Genzel, R. 
1999, ApJ, 517, L13

\bibitem{molinari90}
Molinari, E., Buzzoni, A., \& Chincarini, G.
1990, MNRAS, 246, 576

\bibitem{morganti98}
Morganti, R., Oosterloo, T., \& Tsvetanov, Z. 
1998, AJ, 115, 915

\bibitem{postman95}
Postman, M., \& Lauer, T. R.
1995, ApJ, 440, 28

\bibitem{radecke97}
Radecke, H.-D.
1997, A\&A, 319, 18

\bibitem{sand88a}
Sanders, D. B., Soifer, B. T., Elias, J. H., Madore, B. F.,
Matthews, K., Neugebauer, G., \& Scoville, N. Z. 
1988a, ApJ, 325, 74

\bibitem{sand88b}
Sanders, D. B., Soifer, B. T., Elias, J. H., Neugebauer, G., \& Matthews, K. 
1988b, ApJ, 328, L35

\bibitem{scodeggio98}
Scodeggio, M., Giovanelli, R., \& Haynes, M. P.
1998, AJ, 116, 2728

\bibitem{steiner81}
Steiner, J. E. 
1981, ApJ, 250, 469

\bibitem{storchi95}
Storchi-Bergman, T., Kinney, A. L., \& Challis, P. 
1995, ApJS, 98, 103

\bibitem{turner93}
Turner, T. J., Urry, C. M., \& Mushotsky, R. F.
1993, ApJ, 418, 653

\bibitem{turner97}
Turner, T. J., George, I. M., Nandra, K., \& Mushotsky, R. F.
1997, ApJS, 113, 23

\bibitem{turner98}
Turner, T. J., George, I. M., Nandra, K., \& Mushotsky, R. F.
1998, ApJ, 493, 91

\bibitem{ulvestad84}
Ulvestad, J. S., \& Wilson, A. S. 
1984, ApJ, 285, 439

\bibitem{white92}
White, R. L., \& Becker, R. 
1992, ApJS, 79, 331

\bibitem{xia98a}
Xia, X.-Y., Boller, T., Wu, H., Deng, Z.-G., Gao, Y., Zou, Z.-L., Mao, S.,
\& Borner, G.
1998a, ApJ, 496, L9

\bibitem{xia98b}
Xia, X.-Y., Boller, T., Wu, H., Deng, Z.-G., Gao, Y., Zou, Z.-L., Mao, S.,
\& Borner, G.
1998b, ApJ, 507, L99

\bibitem{xia99}
Xia, X.-Y., Mao, S., Wu, H., Liu, X.-W., Gao, Y., Deng, Z.-G., \& Zou, Z.-L.
1999, preprint (astro-ph/9905202)

\bibitem{yun99}
Yun, M. S., \& Hibbard, J. E. 1999, in preparation

\end{thebibliography}
\end{document}